%% file: Kyzylova_DPF2019_Proceedings.tex
\documentclass[11pt]{article}
\usepackage[margin=1in]{geometry}
\usepackage{graphicx}
\usepackage{amsmath}
\usepackage{lineno}
\input econfmacros.tex 

\def\Title#1{\begin{center} {\Large {\bf #1} } \end{center}}
\def\Author#1{\begin{center} {\normalsize {\sc #1} } \end{center}}
\def\Institution#1{\begin{center} {\normalsize {\it #1} } \end{center}}
\def\Abstract#1{\noindent {\normalsize {\bf Abstract:} {\normalfont #1}}}
\def\Conference{\vspace{4mm}\begin{raggedright} {\normalsize {\it Talk presented at the 2019 Meeting of the Division of Particles and Fields of the American Physical Society (DPF2019), July 29--August 2, 2019, Northeastern University, Boston, C1907293.} } \end{raggedright}\vspace{4mm}}

\begin{document}

\Title{A Search for Sterile Neutrinos with PROSPECT }

\Author{Olga Kyzylova \\ \textit{(for the PROSPECT Collaboration)}}

\Institution{Drexel University, Philadelphia, PA, 19104, USA}

\Abstract{The Precision Reactor Oscillation and Spectrum Experiment (PROSPECT) performs a precision measurement of reactor antineutrinos through inverse beta decay at a baseline range of 7-9 m from the core of the High Flux Isotope Reactor (HFIR). The single, movable detector has a segmented design of 154 optically separated individual segments that serves multiple purposes. Segments, filled with $^{6}$Li-loaded liquid scintillator, cover a range of baselines from the reactor core and allow precise event localization. A reactor-model independent search of eV$^{2}$-scale sterile neutrino oscillations is achieved by performing a relative measurement of the antineutrino event rates and energy distributions between segments within the detector. This talk will discuss the PROSPECT oscillation analysis and present recent results.}

\Conference 

\section{Introduction}

Recent reactor neutrino experiments designed for the study of neutrino oscillations exhibited anomalies in both flux and the antineutrino spectrum. One of the anomalies, the reactor antineutrino anomaly (RAA)~\cite{Ref1}, is a $\sim$6\% average deficit of observed flux with respect to the theoretical predictions~\cite{Ref2, Ref3}. Suggested explanations of this deficit include both flaws in theoretical models of reactor antineutrino spectra and the existence of an eV-scale sterile neutrino state leading to the meter-scale neutrino oscillations. The anomaly in the reactor antineutrino spectrum has an observed excess at 4-7 MeV antineutrino energies in the Daya Bay~\cite{Ref4}, Double Chooz~\cite{Ref5}, and RENO~\cite{Ref6} reactor experiments. This discrepancy indicates the insufficiency of the current reactor models.

The Precision Reactor Oscillation and Spectrum Experiment (PROSPECT) described in this article studies the RAA by looking at antineutrino oscillations at short ($<$12 m) baselines from a highly-enriched $^{235}$U reactor core. The antineutrino flux and spectrum are precisely measured by a detector at a range of baselines from the reactor core.  The goals of the experiment include discovery/exclusion of eV-scale sterile neutrinos, testing reactor antineutrino models with the antineutrino spectrum produced by a highly-enriched uranium reactor core, and obtaining new data complementary to the existing nuclear data measurements.

\section{Experiment Setup}
\begin{figure}[htb]
\centering
\includegraphics[width=0.95\textwidth]{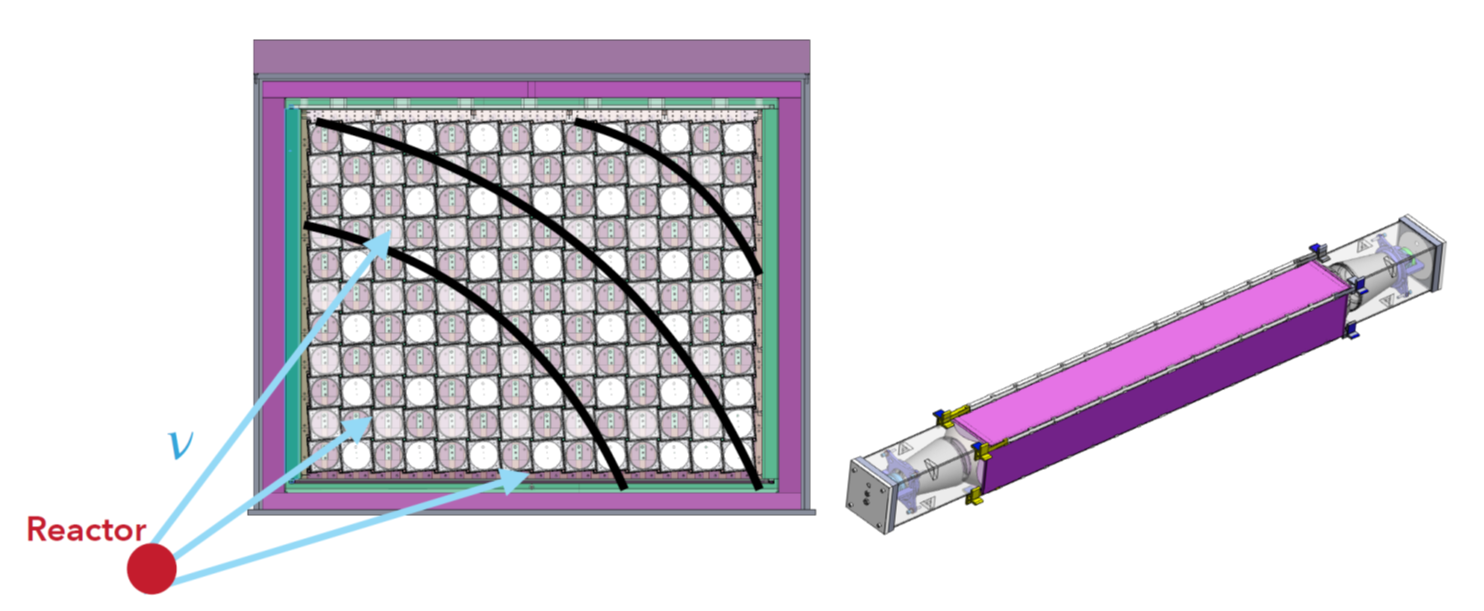}
\caption{Segmented detector design showing how baseline bins are formed (left) and a single optically separated segment (right).}
\label{fig:Detector}
\end{figure}

PROSPECT utilizes the High Flux Isotope Reactor at the Oak Ridge National Laboratory. The detector is based on the inverse beta decay (IBD) reaction with gamma-like prompt and neutron capture-like delayed signals. $^{6}$Li-doped EJ-309-based liquid organic scintillator was chosen for its high light yield and substantial pulse shape discrimination that improves discrimination of the signal from the background~\cite{Ref7}.
The scintillator is segmented into 154 optically separated segments (see Figure \ref{fig:Detector}). Photomultiplier tubes on both ends of each segment provide separate read out. The detector array is shielded with a layered system including lead, borated high-density polyethylene, and water bricks~\cite{Ref8}. The segmentation of the detector with optical panels and hollow support rods allows access to the whole detector volume for \textit{in situ} calibration. Neutron and gamma-ray calibration sources are encapsulated and deployed at different locations inside the detector with tensioned string loops. The sources are used to calibrate energy and neutron detection efficiency in each detector segment. 

\section{Oscillation Analysis}

\begin{figure}[htb]
\centering
\includegraphics[width=0.85\textwidth]{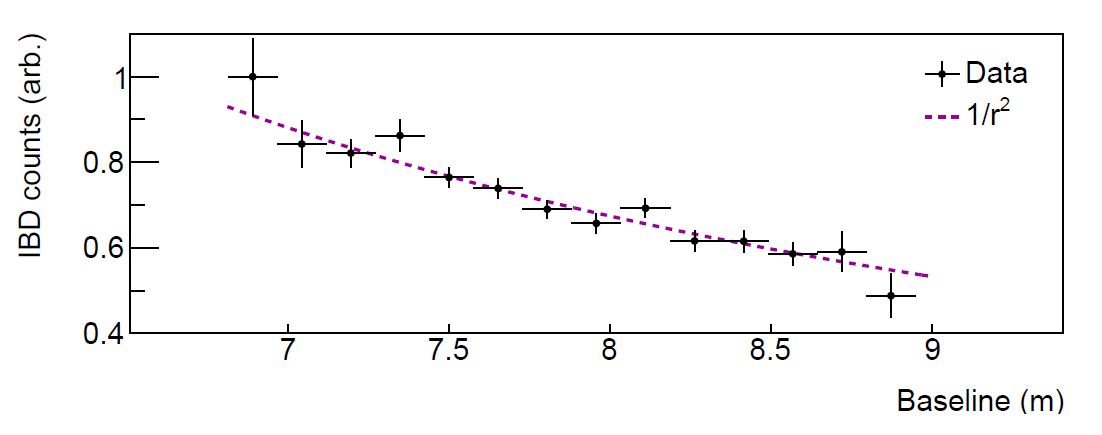}
\caption{Dependence of the rate of IBD events on the distance from the reactor~\cite{Ref9}.}
\label{fig:SpectrumDistance}
\end{figure}

The segmented design of the PROSPECT detector and its proximity to a compact reactor core allow the experiment to cover a wide range of parameter space for the search for sterile neutrino oscillations. Figure \ref{fig:SpectrumDistance} presents the relative IBD rates in the 108 fiducial segments separated into 14 baseline bins as a function of baseline. The rate follows 1/r$^{2}$ behavior throughout the detector volume, as expected, with a 40\% rate decrease from the shortest to the longest baseline of the detector.

PROSPECT performs a reactor model-independent search for sterile neutrinos. Each segment measures the antineutrino spectrum independently, and by separating segments into 6 baseline bins depending on their distance from the reactor, we obtain energy spectra at the respective baselines. To exclude the spectrum shape dependence in the oscillation analysis, we divide the spectrum at each baseline bin by the total spectrum of the full detector fiducial volume normalized to the corresponding segment. Using the formula for the survival probability of electron antineutrinos,
\begin{equation}
    P_{ee}=1-\text{sin}^{2} 2\theta_{14}\cdot\text{sin}^{2}(\text{1.27}\cdot\Delta m^{2}_{41}\frac{L}{E}),
    \label{eq1}
\end{equation}                                      
we compare the relative energy spectra at each baseline to search for sterile neutrino oscillations.

\section{Oscillation Results}
\begin{figure}[htb]
\centering
\includegraphics[width=0.9\textwidth]{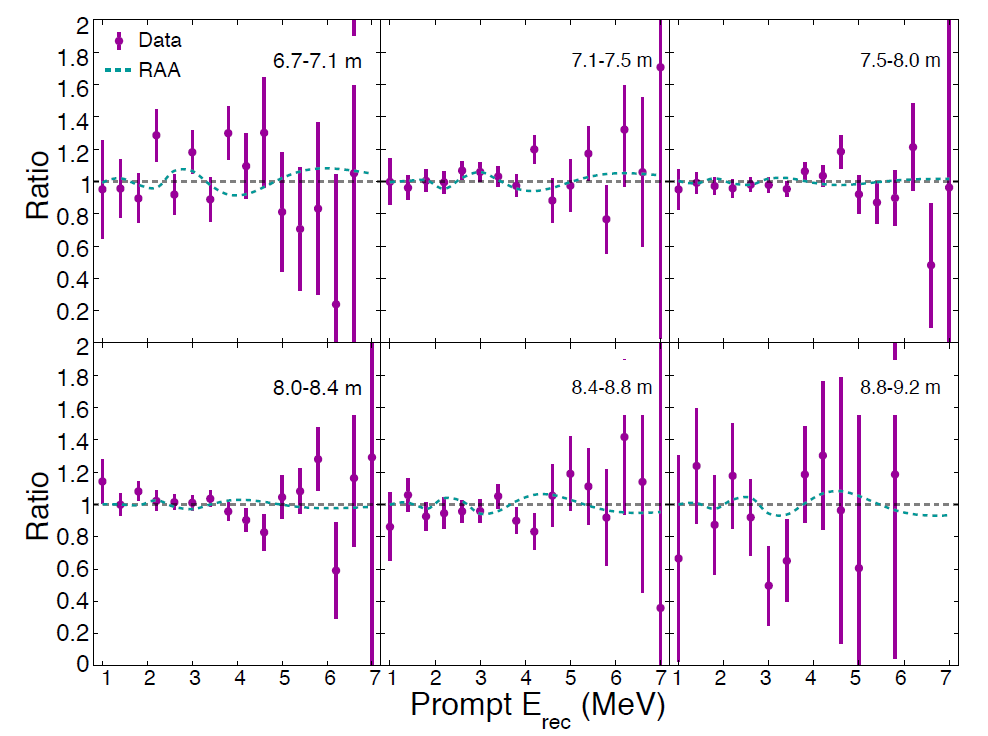}
\caption{Relative IBD spectra for six baseline bins with RAA-best fit point oscillations shown for comparison~\cite{Ref9}.}
\label{fig:SpectrumBaseline}
\end{figure}

A data set comprising 33 days of reactor-on and 28 days of reactor-off was used. During this period, 25461 IBDs were detected with an average of $\sim$771 IBDs/day. A signal-to-background ratio of 1.32 was achieved, which is the best demonstrated signal-to-background ratio for an on-surface reactor experiment.

The results for relative oscillation spectra of 6 baseline bins are shown in Figure \ref{fig:SpectrumBaseline}. Purple points represent data, the green dashed line is the expected oscillation for the RAA-best fit value ($\text{sin}^{2} 2\theta_{14}$ = 0.165, $\Delta m^{2}_{41}$ = 2.39 eV$^{2}$), and the flat black dashed line represents null oscillations. Comparing to the RAA-best fit point, the data do not follow an oscillatory pattern.

To construct the exclusion curve for sterile neutrinos, we define $\chi^{2}$ as
\begin{equation}
\chi^2=\Delta^{T}\text{V}^{-1}_{tot}\Delta.
  \label{eq2}
\end{equation} 

Here, the covariance matrix $\text{V}_{tot}$ is a sum of all covariance matrices $\text{V}_{x}$ produced for each systematic uncertainty and the signal and background statistical uncertainties, and it takes into account the correlation between energy and baseline bins. $\Delta$ is a vector that is defined as the difference between the relative observed and expected spectra in the 6 baseline bins and 16 energy bins:
\begin{equation}
    \Delta_{l,e}=O_{l,e}-O_{e}\frac{E_{l,e}}{E_{e}},
    \label{eq3}
\end{equation}
where $O_{l,e}$ and $E_{l,e}$ are observed and expected rates at $e^{th}$ reconstructed prompt energy bin and $l^{th}$ baseline bin, and $O_{e}$ and $E_{e}$ are corresponding total rates summed over all baselines in $e^{th}$ energy bin:
\begin{equation}
    O_{e}=\sum_{l=1}^{6}O_{l,e},~E_{e}=\sum_{l=1}^{6}E_{l,e}.
    \label{eq4}
\end{equation}

The spectral shape variations here are excluded by dividing the spectra $O_{l,e}$ and $E_{l,e}$ in each baseline bin by the total spectra $O_{e}$ and $E_{e}$ across the whole detector, therefore comparing with equation (1) only the relative differences between observed and predicted spectra. In more details, the procedure is described in~\cite{RefPranavaThesis}.

\begin{figure}[htb]
\centering
\includegraphics[width=0.65\textwidth]{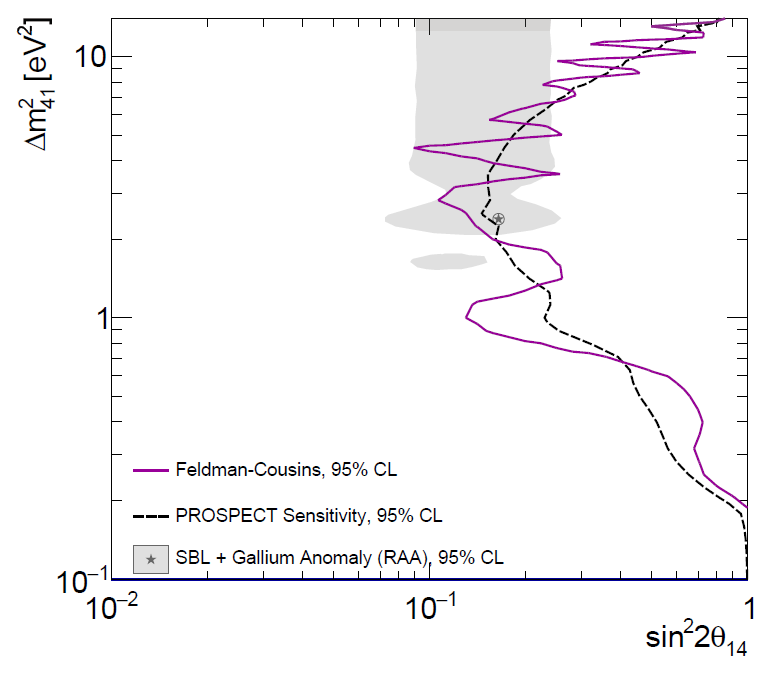}
\caption{PROSPECT exclusion curve and sensitivity at 95\% CL~\cite{Ref9}.}
\label{fig:n4Comparison}
\end{figure}

The final exclusion contour is presented in Figure \ref{fig:n4Comparison}. The Feldman-Cousins approach~\cite{Ref10} was used for construction of confidence interval in order to correctly handle boundary features such as the bounded nature of $\text{sin}^{2}2\theta$ [0, 1] or cases when the oscillation frequency approaches the energy bin size. The PROSPECT exclusion curve for the presented dataset (purple solid line) excludes the RAA-best fit point at $>$ 95\% confidence level. PROSPECT also covers recently reported best-fit point of Neutrino-4 experiment~\cite{Ref11} and disfavors it at $>$ 95\% confidence level.

The importance of using the Feldman-Cousins method for statistical analysis in the search for sterile neutrino oscillations has been recently highlighted by M. Agostini and B. Neumair in~\cite{Ref12}, listing PROSPECT as one of the few experiments that successfully have utilized this method for their analyses.

\section{Conclusions}

In conclusion, PROSPECT performed a search for short-baseline sterile neutrino oscillations from a highly-enriched $^{235}$U reactor. Segmented detector design and relative spectral comparison utilized in the analysis provide a model independent study of sterile neutrino oscillations. With 33 days of reactor-on data, PROSPECT disfavors the RAA sterile neutrino best fit point at 95\%  confidence level (2.2$\sigma$). Oscillation analysis is based on the Feldman-Cousins approach that is necessary to assign correct confidence intervals.

\section*{Acknowledgements}
This material is based upon work supported by the U.S. Department of Energy Office of Science and the Heising-Simons Foundation. Additional support is provided by Illinois Institute of Technology, LLNL, NIST, ORNL, Temple University, and Yale University. We gratefully acknowledge the support and hospitality of the High Flux Isotope Reactor, managed by UT-Battelle for the U.S. Department of Energy.

\bibliography{citations.bib}
 
\end{document}

%% file: econfmacros.tex
\def\beq{\begin{equation}}
\def\eeq#1{\label{#1}\end{equation}}
\def\eeqn{\end{equation}}

\def\beqa{\begin{eqnarray}}
\def\eeqa#1{\label{#1}\end{eqnarray}}
\def\eeqan{\end{eqnarray}}

\let\bar=\overbar

\def\Dslash{\not{\hbox{\kern-4pt $D$}}}
\def\dslash{\not{\hbox{\kern-2pt $\del$}}}

\def\msb{{\bar{\ssstyle M \kern -1pt S}}}

